\documentclass[aps,pra,preprint,groupedaddress,onecolumn]{revtex4-1}
\usepackage{graphicx}

\begin{document}


\title{Kinetic Inductance Magnetometer}


\author{Juho Luomahaara}
\author{Visa Vesterinen}
\author{Leif Gr\"{o}nberg}
\author{Juha Hassel}
\affiliation{VTT Technical Research Center of Finland, Tietotie 3, 02150 Espoo, Finland}


\date{\today}

\begin{abstract}
Ultrasensitive magnetic field detection is utilized in the fields of science, medicine and industry. We report on a novel magnetometer relying on the kinetic inductance of superconducting material. The kinetic inductance exhibits a non-linear response with respect to DC current, a fact that is exploited by applying magnetic flux through a superconducting loop to generate a shielding current and a change in the inductance of the loop. The magnetometer is arranged into a resonator, allowing readout through a transmission measurement that makes the device compatible with radio frequency multiplexing techniques. The device is fabricated using a single thin-film layer of NbN, simplifying the fabrication process compared to existing magnetometer technologies considerably. Our experimental data, supported by theory, demonstrates a magnetometer having potential to replace established technology in applications requiring ultra-high sensitivity. 
\end{abstract}

\maketitle


Sensing ultra-low magnetic fields has various applications, such as magnetoencephalography (MEG) \cite{ham}, magnetocardiography (MCG) \cite{sek}, ultra-low-field nuclear magnetic resonance \cite{mcd2} and magnetic resonance imaging (ULF MRI) \cite{mcd,zot,ves}, exploration of magnetic minerals \cite{chw} and a wide range of other scientific purposes. The most established method has been to use low critical temperature superconducting quantum interference devices (SQUIDs) \cite{cla,ryh} featuring field sensitivity in the fT/Hz$^{1/2}$ regime or below, although also other techniques exist \cite{fal,pan,kom,gia}. We describe a new magnetometer based on the current non-linearity of superconducting material. The non-linearity stems from kinetic inductance, which during the last decade has been actively pursued in the fields of submillimeter-wave detection \cite{day,tim} and parametric amplification \cite{eom}. Different versions of magnetometers based on kinetic inductance have been reported previously \cite{mes,aye,khe}. Yet, the best reported field sensitivities have been in the order of pT/Hz$^{1/2}$ \cite{aye}. The benefit of our approach is extreme simplicity: the device is fabricated from a single layer of niobium nitride. Furthermore, radio frequency multiplexing techniques \cite{mch} can be applied, enabling the simultaneous readout of multiple sensors, which is essential, e.g., in biomagnetic measurements requiring data from large sensor arrays \cite{sar}. We demonstrate a device achieving field sensitivity in the fT/Hz$^{1/2}$ range and a wide dynamic range without feedback.

Consider a superonducting loop composed of thin film with thickness $h$ and linewidth $w$. When $h\ll\lambda$, the current density $J_{\mathrm{s}}$ can be assumed to be homogeneous within the cross-section and the magnetic flux quantization through the ring reads
\begin{equation}
	(L_{\mathrm{k}}+L_{\mathrm{g}})I_{\mathrm{s}}-\Phi_{\mathrm{a}}=m\Phi_{0},
\end{equation}
where $\lambda$ is the magnetic penetration depth, $I_{\mathrm{s}}$=$J_{\mathrm{s}}wh$ the shielding current, $\Phi_{\mathrm{a}}$ the applied magnetic flux, $m$ an integer and $\Phi_{0}$ = 2.07 fWb the flux quantum. Unlike the geometric inductance $L_{\mathrm{g}}$ that is associated with the magnetic field of the ring, the kinetic inductance $L_{\mathrm{k}}$=$\mu_{0}\lambda^{2}l/wh$ ($\mu_{0}$ is the vacuum permeability and $l$ the length of the loop) stems from the motion of the Cooper pairs. Kinetic inductance becomes non-linear with respect to shielding current $I_{\mathrm{s}}$ as the kinetic energy approaches the depairing energy of the Cooper pairs \cite{eom,ann}. With sufficiently small $I_{\mathrm{s}}$, the non-linearity assumes a form
\begin{equation}
	L_{\mathrm{k}}(I_{\mathrm{s}})=L_{\mathrm{k0}}\left(1+\left(\frac{I_{\mathrm{s}}}{I^{*}}\right)^2\right).
\end{equation}
Here, $L_{\mathrm{k0}}$ is the kinetic inductance at zero current and $I^{*}$, a parameter usually of the order critical current $I_{\mathrm{c}}$, sets the scale for the non-linearity.

The magnetic flux $\Phi_{\mathrm{a}}$ applied through the loop (inductance $L=L_{\mathrm{g}}+L_{\mathrm{k}}$) generates a shielding current $I_{\mathrm{s}}$ according to Equation (1), which in turn modifies the inductance of the loop through Equation (2), revealing the operation principle behind the kinetic inductance magnetometer. Figure 1\textbf{a} shows a practical realization of such a device in planar geometry along with the simplified  measurement setup depicted in Figure 1\textbf{b}. A square-shaped superconducting loop (width $W$ = 20 mm) has been fabricated from thin film NbN (critical temperature $T_{\mathrm{C}}$ = 14 K). An interdigital capacitor $C$ is arranged in parallel with the loop, and the formed resonator is further coupled to a transmission line (characteristic impedance $Z_{0}$ = 50 $\Omega$) with a matching capacitor $C_{\mathrm{c}}$. The inductance change is now read out by measuring the transmission $S_{21}$ through the resonator. Below we quantify the magnetic bias and excitation as an average orthogonal field $B_{0} = \Phi_\mathrm{a}/A$ with $A$ the surface area of the loop.
\begin{figure}[htpb]
\begin{center}
\includegraphics[width=13cm]{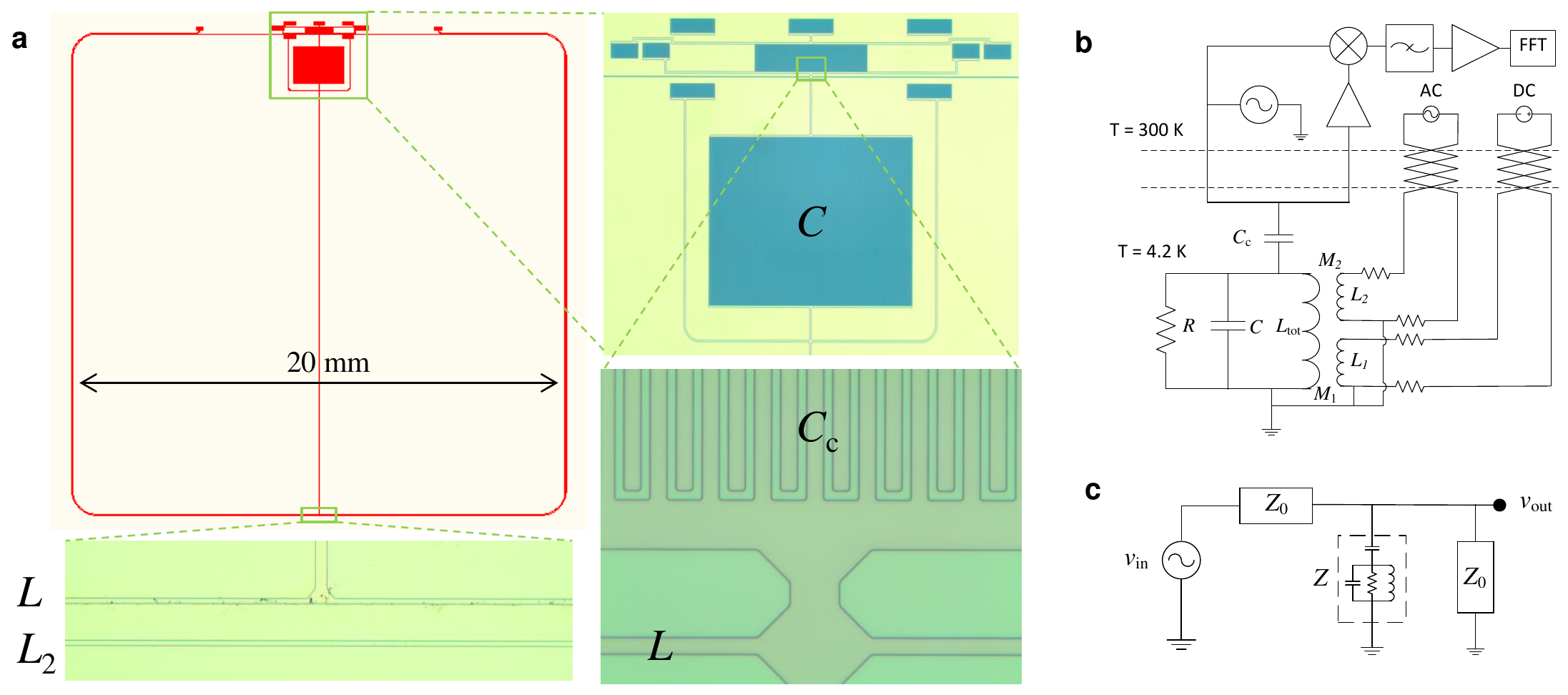}
\caption{Magnetometer design and simplified circuit diagrams for the setup and the transmission measurement. \textbf{a}, The device contains a superconducting loop $L$ fabricated employing a single NbN thin film layer. The interdigital capacitors $C$ and $C_{\mathrm{c}}$, shown in the optical microphotograph zoom-ins, are placed inside and outside the loop, respectively. Together with the inductive loop, the capacitors form a resonator coupled to an external transmission line. \textbf{b}, An off-chip coil $L_{1}$ and a superconducting on-chip coil $L_{2}$ with calibrated mutual inductances ($M_{1}$ and $M_{2}$) are arranged around the magnetometer loop to generate bias and excitation fluxes. An RF generator supplies excitation and reference signals for the resonator and mixer, respectively. The applied magnetic field changes the kinetic inductance of the loop affecting the transmission spectrum of the microwaves passing the resonator. The resulting signal, being proportional to the excitation voltage $v_{\mathrm{in}}$, is amplified, mixed down to DC and finally Fourier transformed to frequency domain. \textbf{c}, Transmission $S_{21}$ can be modelled with an equivalent circuit, which yields $S_{21}=2v_{\mathrm{out}}/v_{\mathrm{in}}=2Z/(2Z+Z_{0})$.} 
\end{center}
\end{figure}

The impedance $Z$ of the capacitively coupled resonator near the resonance frequency $\omega$ $\approx$ $\omega_{0}$ = $1/\sqrt{L_{\mathrm{tot}}(C+C_{\mathrm{c}})}$ can be given in the form 
\begin{equation}
	Z \approx \frac{Z_{0}Q_{\mathrm{ext}}}{2Q_{\mathrm{i}}}(1+j2Q_{\mathrm{i}}(\omega-\omega_{0})) 
\end{equation}
assuming $Q_{\mathrm{i}}\gg1$. Here $Q_{\mathrm{i}}$ is the intrinsic quality factor describing the superconductor losses, $Q_{\mathrm{ext}}=2(C+C_{\mathrm{c}})/(\omega_{0}Z_{0}C_{\mathrm{c}}^2)$ the coupling quality factor and the inductance $L_{\mathrm{tot}} = L/4+L_{\mathrm{par}}$ contains the contributions from the loop inductance and parasitics. Differentiating Equations (1)-(3), enables the calculation of the device responsitivity $\frac{\partial v_{\mathrm{out}}}{\partial B}$, where $v_{\mathrm{out}}$ is the output voltage of the transmission measurement (see Figure 1\textbf{c}). The responsitivity is an imaginary quantity at resonance and can be approximated as  
\begin{equation}
\left|\frac{\partial v_{\mathrm{out}}}{\partial B}\right|_{\omega = \omega_0} = \left|\frac{\partial v_{\mathrm{out}}}{\partial L_{\mathrm{k}}}\frac{\partial L_{\mathrm{k}}}{\partial I_{\mathrm{s}}}\frac{\partial I_{\mathrm{s}}}{\partial \Phi_{\mathrm{a}}}A\right|_{\omega = \omega_0} = \frac{v_{\mathrm{in}}Q_{\mathrm{t}}^{2}}{4Q_{\mathrm{ext}}L_{\mathrm{tot}}}\frac{I_{\mathrm{s}}}{I^{*2}\left(1+3\left(\frac{I_{\mathrm{s}}}{I^{*}}\right)^2+\frac{L_{\mathrm{g}}}{L_{\mathrm{k}0}}\right)}A,
\end{equation}
where the resonator is excited with a voltage $v_{\mathrm{in}}$ and the total quality factor $Q_{\mathrm{t}}$ along with the inductance $L_{\mathrm{tot}}$ are functions of the bias current $I_{\mathrm{s}}$. 

Ultimately, the device gain is limited by the excitation $v_{\mathrm{in}}$ generating an RF current (amplitude $i_{\mathrm{L}}$) through inductance $L_{\mathrm{tot}}$. In the limit $i_{\mathrm{L}}=2(I_{\mathrm{c}}-I_{\mathrm{s}})$, where the factor of two stems from the loop geometry, the magnetometer is driven to the normal state and the maximum gain for a given bias current $I_{\mathrm{s}}$ becomes 
\begin{equation}
	\left|\frac{\partial v_{\mathrm{out}}}{\partial B}\right|_{\omega=\omega_{0}}^{\mathrm{max}} \approx \frac{Z_{0}\omega_{0}^{2}Q_{\mathrm{i}}C_{\mathrm{c}}C}{2C+Z_{0}\omega_{0}Q_{\mathrm{i}}C_{\mathrm{c}}^{2}}\frac{I_{\mathrm{s}}(I_{\mathrm{c}}-I_{\mathrm{s}})}{I^{*2}\left(1+3\left(\frac{I_{\mathrm{s}}}{I^{*}}\right)^2+\frac{L_{\mathrm{g}}}{L_{\mathrm{k}0}}\right)}A.
\end{equation}

The fluctuation mechanisms peculiar to kinetic inductance devices are related to the dynamics of quasiparticle excitations: the thermal motion of the unpaired electrons and the stochastics of the quasiparticle density. The former, combined with the contribution from thermal fluctuations of the feeding line, generates a Johnson voltage noise $S_{\mathrm{v}}^{1/2}=\sqrt {4k_{\mathrm{B}}TZZ_{0}/(Z+Z_{0})}$ at temperature $T$, where $k_{\mathrm{B}}$ is the Boltzmann constant. Taking into account the coupling to the input of the amplifier and utilizing the Equation (4), this can be mapped to magnetic field noise at the magnetometer loop
\begin{equation}
	S_{B,\mathrm{th}}^{1/2} = \sqrt{8k_{\mathrm{B}}TZ_{0}Q_{\mathrm{ext}}(2Q_{\mathrm{i}}+Q_{\mathrm{ext}})}\frac{Q_{\mathrm{i}}+Q_{\mathrm{ext}}}{Q_{\mathrm{ext}}Q_{\mathrm{i}}^{2}}\left(1+3\left(\frac{I_{\mathrm{s}}}{I^{*}}\right)^2+\frac{L_{\mathrm{g}}}{L_{\mathrm{k}0}}\right)\frac{L_{\mathrm{tot}}I^{*2}}{I_{\mathrm{s}}v_{\mathrm{in}}A}.
\end{equation}
The generation-recombination of the charge carriers, on the other hand, leads them to fluctuate according to spectral density $S_{\mathrm{qp}}(\omega)=4N_{\mathrm{qp}}\tau_{\mathrm{r}}/(1+(\omega\tau_{\mathrm{r}})^2)$, where $N_{\mathrm{qp}}$ = $Vn_{\mathrm{qp}}$ is the number of quasiparticles in the superconductor volume $V$ and $\tau_{\mathrm{r}}$ the quasiparticle recombination time \cite{vis}. The field noise due to the generation-recombination process can be evaluated in the low-frequency limit $\omega\ll1/\tau_{\mathrm{r}}$ as (see Supplement material)
\begin{equation}
	S_{B,\mathrm{gr}}^{1/2} = 2L_{\mathrm{k}0}\sqrt{\frac{n_{\mathrm{qp}}\tau_{\mathrm{r}}}{Vn_{\mathrm{s}}^{2}}}\left(\frac{\partial L}{\partial B}\right)^{-1},
\end{equation}
where $n_{\mathrm{s}}$ and $n_{\mathrm{qp}}$ are the densities of Cooper pairs and quasiparticles, respectively.

\begin{figure}[htpb]
\begin{center}
\includegraphics[width=13cm]{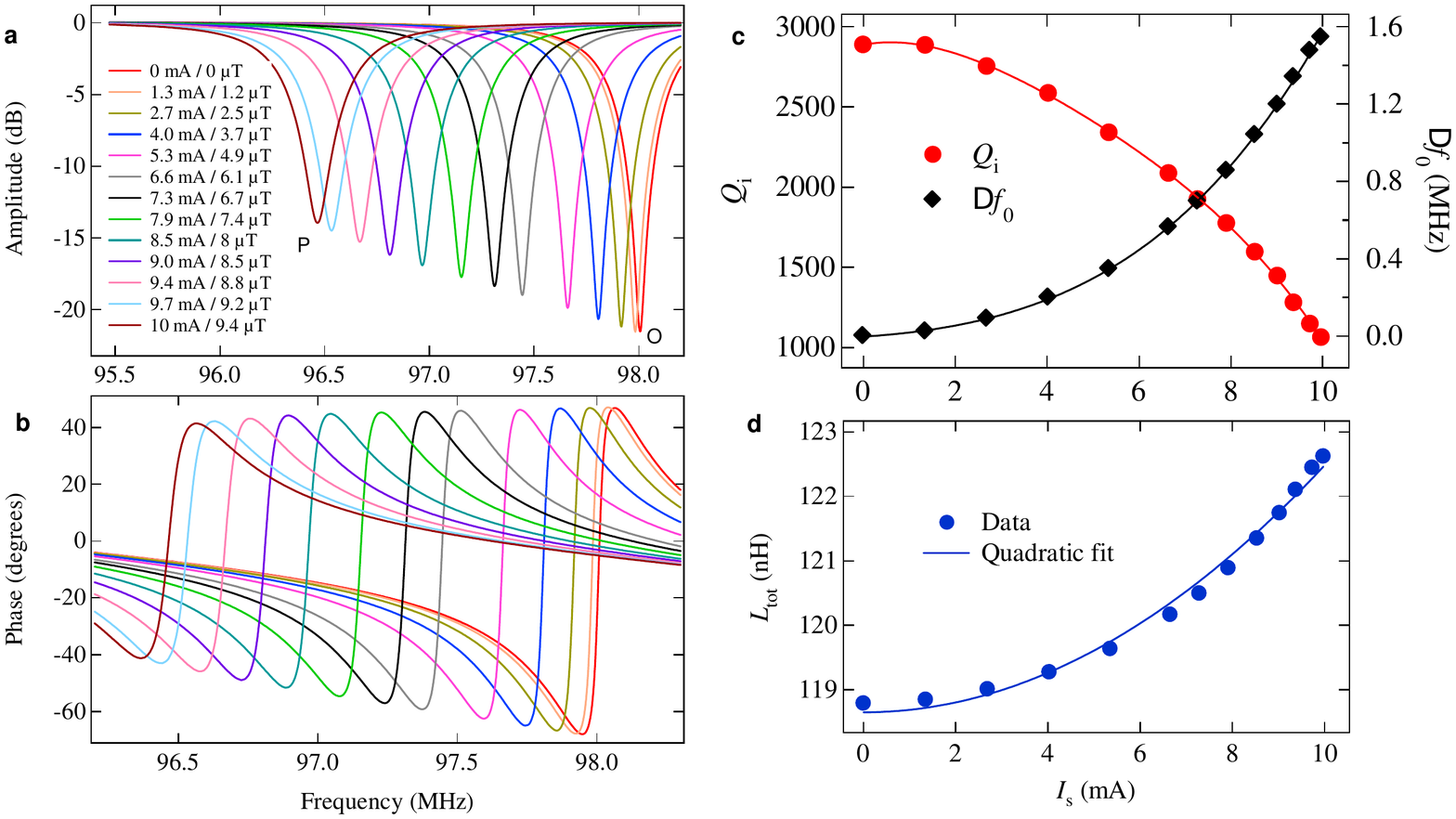}
\caption{Transmission spectra and derived device parameters. \textbf{a,b}, Resonator transmission $S_{21}$ was measured with a network analyzer as a function of the screening current $I_{\mathrm{s}}$ and bias field $B_{0}$. \textbf{c}, The intrinsic quality factor $Q_{\mathrm{i}}$ = $R(I_{\mathrm{s}})\sqrt{(C+C_{\mathrm{c}})/L_{\mathrm{tot}}(I_{\mathrm{s}})}$ of the resonator and the resonance frequency shift $\Delta f_{0}=f_{0}(I_{\mathrm{s}}=0)-f_{0}(I_{\mathrm{s}})$ have been obtained from transmission curves of \textbf{a} and \textbf{b} (see Supplement material). Solid lines present polynomial fits. \textbf{d}, The inductance $L_{\mathrm{tot}}$ determines the resonance frequency according to $f_{0}(I_{\mathrm{s}})$= $1/(2\pi\sqrt{L_{\mathrm{tot}}(I_{\mathrm{s}})(C+C_{\mathrm{c}})}$, where $C$ = 20 pF and $C_{\mathrm{c}}$ = 2.2 pF. The inductance $\mathrm{L_{\mathrm{tot}}}=(L_{\mathrm{g}}+L_{\mathrm{k}})/4+L_{\mathrm{par}}$ contains contributions from loop inductance and parasitics, where $L_{\mathrm{g}}$ = 142 nH and $L_{\mathrm{par}}$ = 28 nH have been computed with a simulation software. The quadratic fit of \textbf{d} gives $L_{\mathrm{k}0}$ = 220 nH and $I^{*}$ = 38 mA for the kinetic inductance.} 
\end{center}
\end{figure}
As discovered from Equation (3), the impedance $Z$ of the resonator reaches a real value $Z_{\mathrm{min}} \approx Z_{0}Q_{\mathrm{ext}}/(2Q_{\mathrm{i}})$ at the resonance $f_{0} = 1/(2\pi\sqrt{L_{\mathrm{tot}}(I_{\mathrm{s}})(C+C_{\mathrm{c}})})$ creating a decline in transmission $S_{21}$ depicted in Figure 2\textbf{a}. As the applied field $B_{0}$, and consequently $I_{\mathrm{s}}$ (assuming $m$ = 0), is adjusted between zero and a maximum value, the resonance dip $f_{0}$ shifts between points O and P, respectively. When $I_{\mathrm{s}}$ is driven above $I_{\mathrm{c}}$ = 10 mA (point P), the integer $m$ assumes a new value and the current $I_{\mathrm{s}}$ resets according to Equation (1). For most of our devices, $I_{\mathrm{s}}$ drops close to zero after exceeding the critical current and the resonance dip hops back from state P to state O. As the external field is changed further, the resonance frequency $f_{0}$ starts again the approach towards point P.

The decrease of Cooper pair density with growing $I_{\mathrm{s}}$ degrades the intrinsic quality factor $Q_{\mathrm{i}}$ of the device and enhances the resonator inductance $L_{\mathrm{tot}}$ due to the kinetic term given by Equation (2). Figures 2\textbf{c} and \textbf{d} show these quantities along with the resonance frequency shift $\Delta f_{0}$. The inductance change enables the derivation of parameters $L_{\mathrm{k}0}$ = 220 nH and $I^{*}$ = 38 mA. Assuming the nominal geometry of the superconducting strip ($l$ = 80 mm, $w$ = 5 $\mu$m and $h$ = 165 nm), one can determine the penetration depth $\lambda$ = 1.3 $\mu$m for the NbN thin film. The value is close to the zero-temperature magnetic penetration depth in the impure limit $\lambda_{0}=\sqrt{\hbar\rho/(\mu_{0}\pi\Delta)}$ = 1.2 $\mu$m, where $\hbar=h/(2\pi)$ is the Planck constant, $\rho$ = 690 $\mu\Omega$ cm the film resisitivity at room temperature and $\Delta$ = 2.5 meV the gap energy at zero temperature for NbN \cite{kam}.

\begin{figure}[htpb]
\begin{center}
\includegraphics[width=7cm]{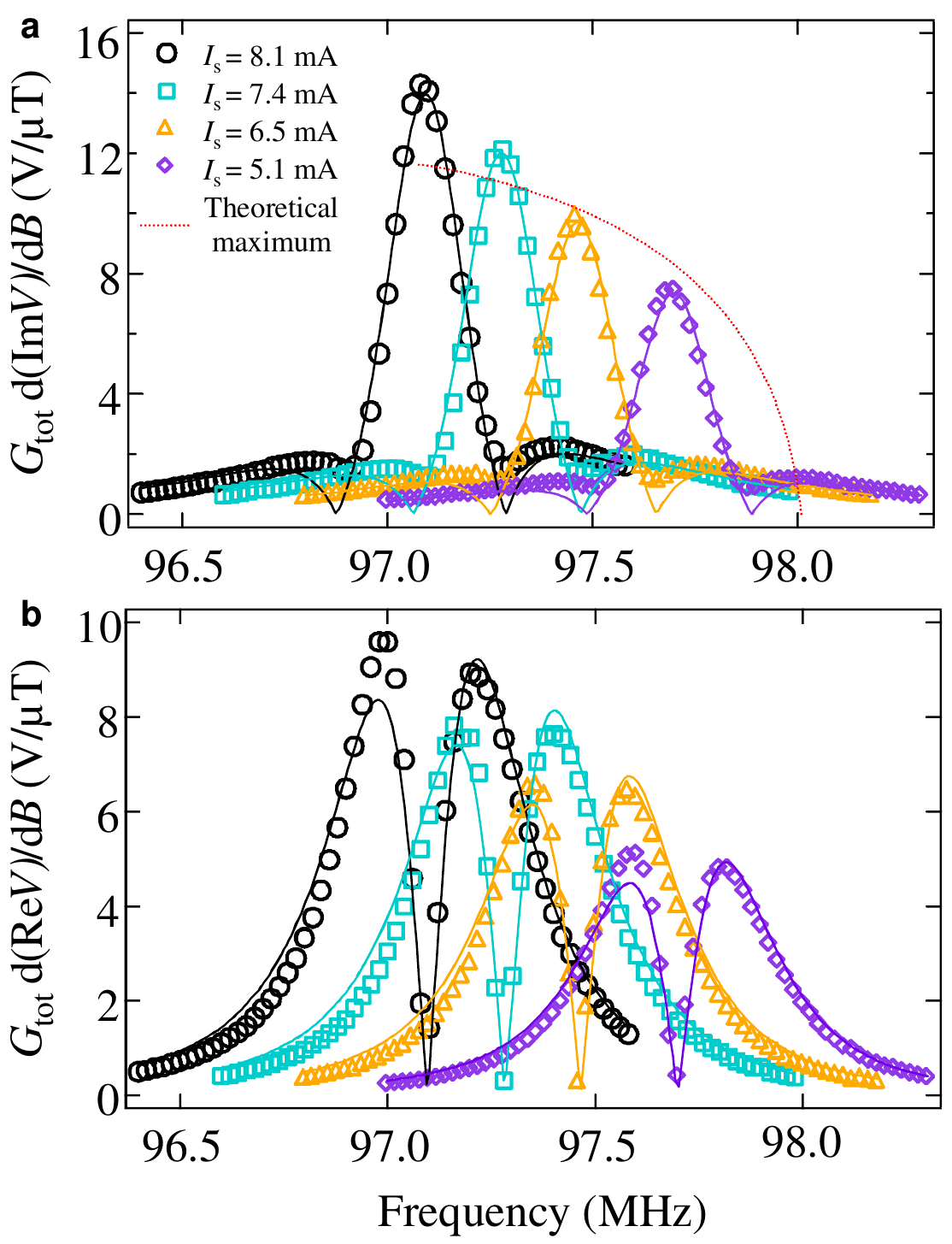}
\caption{Device gain. \textbf{a,b}, The measured device responsitivities $G_{\mathrm{tot}}d(\mathrm{Im} V)/dB$ and $G_{\mathrm{tot}}d(\mathrm{Re} V)/dB$ representing  quadrature and in-phase components, respectively, are shown along with theoretical fits (solid lines), where the total voltage gain $G_{\mathrm{tot}}$ and bias current $I_{\mathrm{s}}$ were chosen as fitting parameters (see Supplement material). The total gain $G_{\mathrm{tot}}$ = $GL_{\mathrm{mix}}L_{\mathrm{att}}$ includes the contributions from the preamplification $G$, and the losses due to mixing $L_{\mathrm{mix}}$ and the attenuation $L_{\mathrm{att}}$ present in the setup. The Equation (4) scaled with $G_{\mathrm{tot}}$ = 1250 gives an estimate for the Q-component at resonance (red dashed line), in reasonable agreement with the measurements.}  
\end{center}
\end{figure}
The measured device responsitivities $G_{\mathrm{tot}}d(\mathrm{Im}V)/dB$ and $G_{\mathrm{tot}}d(\mathrm{Re}V)/dB$ are depicted in Figure 3\textbf{a} and \textbf{b} for quadrature (Q) and in-phase (I) components, respectively. The theoretical plots, using parameters derived from the transmission measurements, follow the measured data yielding $G_{\mathrm{tot}}$ = 1250$\pm$60 for the system gain, which is in line with the nominal value $G_{\mathrm{nom}}$ = 1120. For a constant excitation $v_{\mathrm{in}}$ = 8.8 mV, the current $i_{\mathrm{L}}$ is a weak function of $I_{\mathrm{s}}$ and can be approximated as $i_{\mathrm{L}}$ $\approx$ 2.1 mA. In this case, flux trapping was observed below the resonance frequency $f_{0}\approx$ 97 MHz corresponding to bias current of the order $I_{\mathrm{s}} \approx$ 8.4 mA, approximately fulfilling the maximum gain condition $i_{\mathrm{L}}$ $\sim$ 2($I_{\mathrm{c}}-I_{\mathrm{s}}$).

The equivalent magnetic field noise $S_{B}$ of the detector was measured by operating the device without external magnetic bias using persistent current  $I_{\mathrm{s}} = m\Phi_{0}/(L_{\mathrm{g}}+L_{\mathrm{k}})$ with finite $m$ to determine the operating point. The RF frequency and amplitude were set to approximately maximize the field responsivity $G_{\mathrm{tot}}dV/dB$. The noise spectrum $S_B^{1/2} = S_{V}^{1/2}/(G_{\mathrm{tot}}dV/dB)$, where $S_{V}^{1/2}$ is the voltage noise at the output, is shown in Figure 4, yielding white noise of about 23$\pm$2 fT/Hz$^{1/2}$ with the error limit stemming from the uncertainty in mutual inductance $M_{1}$ calibration. Two electronics related fluctuation mechanisms affecting the result were identified. One such mechanism is the white background noise originating from readout amplifier and the excitation source (see the Supplement for details). The contribution of these sources was determined by choosing a bias frequency slightly off-resonance and recording the background output noise level corresponding to an equivalent noise of 17 fT/Hz$^{1/2}$. Another noise generating mechanism is the RF amplitude fluctuations, manifesting itself through current rectification due to non-linear inductance. An estimate of about 10 fT/Hz$^{1/2}$ was obtained for this contribution through a separate characterization of the RF source. Thus, the electronics contribution is altogether $\sim$ 20 fT/Hz$^{1/2}$ that, within the uncertainties, approximately explains the measured noise level. The thermal noise of the device, as computed from Equation (6), gives $S_{\mathrm{B,th}}^{1/2}$ = 2 fT/Hz$^{1/2}$. Furthermore, the generation-recombination noise from Equation (7) was estimated to be negligible, i.e., 0.03 fT/Hz$^{1/2}$. Therefore, by improving the electronics one should be able to reach a field sensitivity comparable with SQUID based magnetometers, even with our first non-optimized devices.
\begin{figure}[htpb]
\begin{center}
\includegraphics[width=8cm]{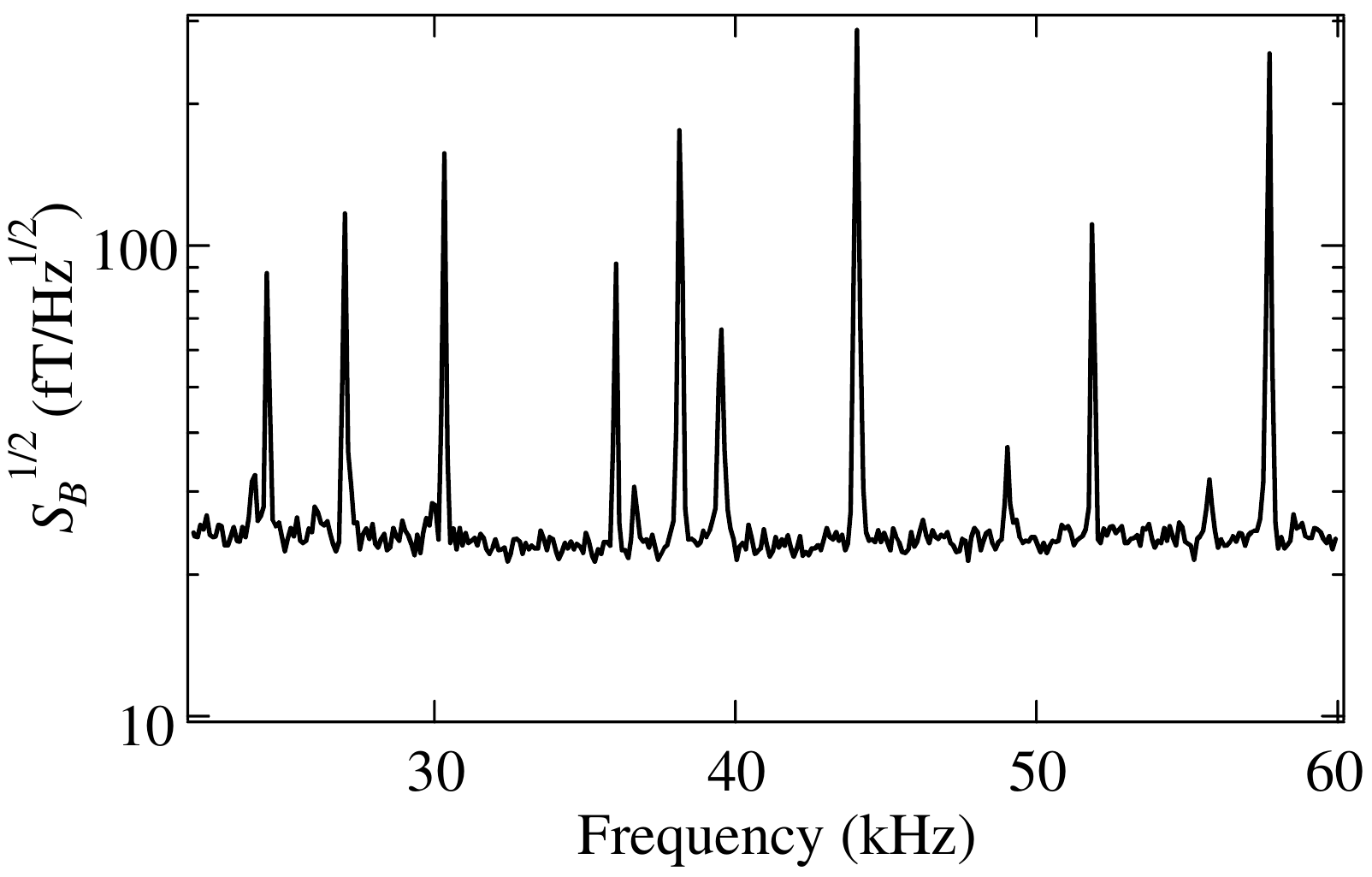}
\caption{System noise. The measured spectral density of the equivalent magnetic field noise $S_{B}^{1/2}$ = 23 fT/Hz$^{1/2}$.}  
\end{center}
\end{figure}

In summary, a magnetometer based on non-linear kinetic inductance with the field sensitivity approaching fT/Hz$^{1/2}$ level was demonstrated. Simple fabrication and multiplexability make it a potential candidate for applications requiring high-sensitivity magnetic field detection. Additional benefits include a high dynamic range and tolerance against ambient magnetic fields. For example, the periodic response of SQUID amplifiers operating without feedback typically limits the linear range below 1 nT. For the devices reported here, the dynamic range is several hundreds of nanotesla. Furthermore, the devices can fundamentally operate without shielding in higher ambient magnetic fields, in contrast to SQUIDs for which the limiting factor is the magnetic penetration through the Josephson junction suppressing responsivity at fields of the order of $\Phi_0 / (\lambda d)$, where $d$ is the dimension of the Josephson junction. This may beneficial for techniques requiring magnetic manipulation of the sample, such as ULF MRI.

\section*{Acknowledgements}
The authors wish to thank Andrey Timofeev, Panu Helist{\"o}, Mikko Kiviranta and Arttu Luukanen for useful discussions as well as Paula Holmlund and Harri Pohjonen for the help in sample preparation. The work was financially supported by the Academy of Finland through Centre of Excellence in Low Temperature Quantum Phenomena and Devices.

\pagebreak

\section*{Supplementary material}
\subsection*{General}The experiments were performed in liquid helium using a cryoprobe equipped with two coaxial rf lines for the transmission measurement of the resonator as well as two dc lines for magnetic field generation. A magnetic shield made of Pb and cryoperm was mounted around the sample resulting in ambient noise level below 5 fT/Hz$^{1/2}$ in the frequency region of interest verified with a SQUID magnetometer. 

\subsection*{Fabrication} The NbN films were formed by reactive sputtering of niobium in nitrogen atmosphere on thermally oxidised silicon substrates. The devices were patterned through optical lithography and reactive ion etching. The critical temperature and resistivity of the film were verified by measuring the resistance as a function of temperature from reference structures on the same wafer as the magnetometers.

\subsection*{Transmission measurements} The field dependency of the device shown in Figure 2 was characterized in transmission measurements. The bias field was varied by tuning the DC voltage supplied through resistors $2R_1$ = 300 $\Omega$ and recording the corresponding transmission spectra with a network analyzer. The spectra were then used for fitting the transmission $S_{21}=2v_{\mathrm{out}}/v_{\mathrm{in}}=2Z/(2Z+Z_{0})$ (see Figure 1\textbf{c}), where the complete impedance of the capacitively coupled resonator 
\begin{equation}
	Z=\frac{1}{j\omega C_{\mathrm{c}}}+\frac{RL_{\mathrm{tot}}/C}{RL_{\mathrm{tot}}+R/(j\omega C)+L_{\mathrm{tot}}/C}
\end{equation}
was employd. Capacitor values, measured with an impedance analyser from similar test structures, were assumed for $C_{\mathrm{c}}$ = 2.2 pF and $C$ = 20 pF, while resonance frequency $f_{0}$ = $1/(2\pi\sqrt{L_{\mathrm{tot}}(I_{\mathrm{s}})(C+C_{\mathrm{c}})}$ and quality factor $Q_{\mathrm{i}}$ = $R(I_{\mathrm{s}})\sqrt{(C+C_{\mathrm{c}})/L_{\mathrm{tot}}(I_{\mathrm{s}})}$ were used as fitting parameters. The amplitude of the transmission were fixed to unity sufficiently far away from the resonance. The best fit was achieved with $Z_{0}$ = 55 $\Omega$, especially a bit off from resonance. As a result, the curves of Figure 2\textbf{c} were obtained. Solid lines represent polynomial fits containing all the necessary information for reproducing the transmission curves of Figure 2\textbf{a} and \textbf{b}.

\begin{figure}[htpb]
\begin{center}
\includegraphics[width=13cm]{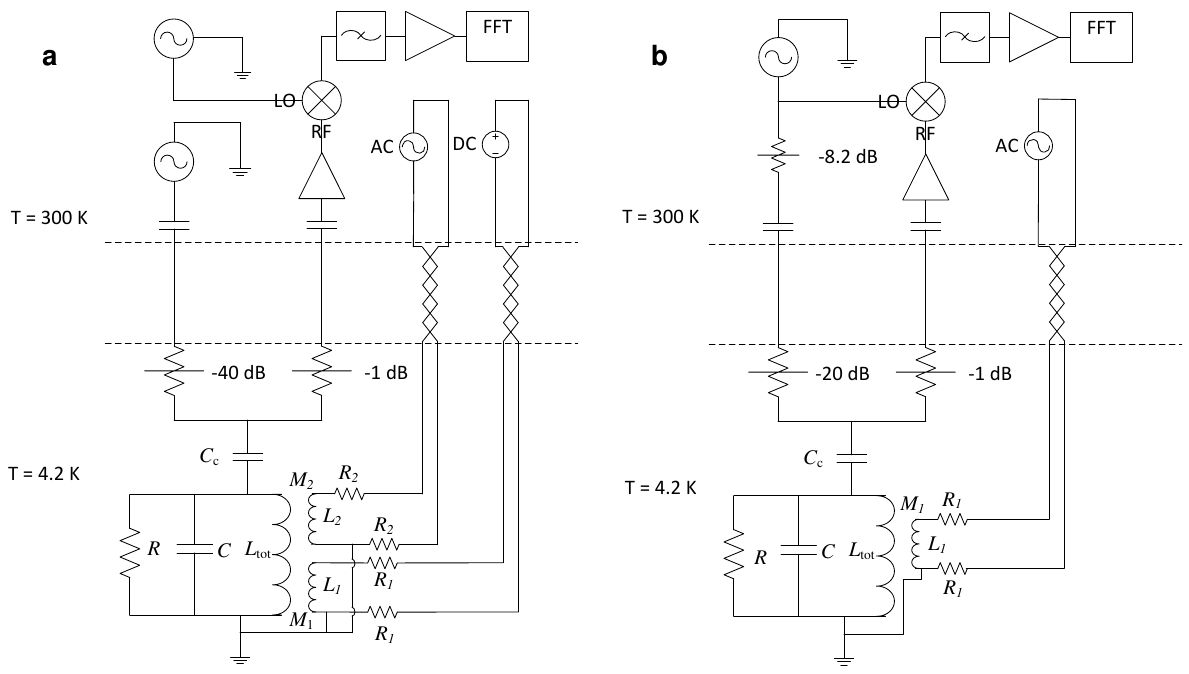}
\caption{Setups used in \textbf{a} responsitivity and \textbf{b} noise measurements.}  
\end{center}
\end{figure}
\subsection*{Responsitivity measurements} The device responsivities $G_{\mathrm{tot}}dV/dB$ were measured with the setup of Figure 5\textbf{a}. RF generators provide signals for excitation $v_{\mathrm{RF}}$ = $v_{0,\mathrm{RF}}\sin(\omega_{\mathrm{RF}}t)$ of the resonator as well for a reference $v_{\mathrm{LO}}$ = $v_{0,\mathrm{LO}}\sin(\omega_{\mathrm{LO}}t+\phi)$ of the mixer. The transmitted signal is amplified at room temperature with a low-noise amplifier and mixed down to DC ($\omega_{\mathrm{LO}} = \omega_{\mathrm{RF}}$). The DC signal is low-pass filtered, amplified and, finally, Fourier transformed to frequency domain. An attenuator (-40 dB) was placed at low temperature end of the input to reduce the noise originating from the room-temperature electronics and wiring. 

The phase $\phi$ of the local oscillator was first tuned to a value $\phi$ = $\phi_{\mathrm{max}}$ providing a maximum response at a given resonance frequency $f_{0}$. At this operating point, the mixing yields a quadrature (Q) component of the signal for $\phi$ = $\phi_{\mathrm{max}}$ and an in-phase (I) component for  $\phi$ = $\phi_{\mathrm{max}}$-$\pi$/2. Using this strategy, the data shown in Figure 3 was measured. The RMS value of the carrier signal $v_{0,\mathrm{RF}}/\sqrt{2}$ was set to 0.88 V corresponding to $v_{\mathrm{in}}$ = $v_{0,\mathrm{RF}}$/(100$\sqrt{2}$) = 8.8 mV after the attenuator. Fitting was performed to measured data using functions of the form $G_{\mathrm{tot}}v_{\mathrm{in}}d(\mathrm{Im}(S_{21}))/dB$ (Q-component) and $G_{\mathrm{tot}}v_{\mathrm{in}}d(\mathrm{Re}(S_{21}))/dB$ (I-component) with $G_{\mathrm{tot}}$ and $I_\mathrm{s}$ selected as fitting parameters. Here transmission $S_{21}$ was computed using the fits of Figure 2\textbf{c} and the total voltage gain $G_{\mathrm{tot}}$ = $GL_{\mathrm{mix}}L_{\mathrm{att}}$ includes contributions from the amplifiers $G$, and the losses due to mixing $L_{\mathrm{mix}}$ and other attenuation $L_{\mathrm{att}}$ present in the setup. The bias current $I_{\mathrm{s}}$ values for each plot are shown in the inset of Figure 3\textbf{a} and the total gain $G_{\mathrm{tot}}$ = 1250$\pm$60, which is close to the nominal gain $G_{\mathrm{nom}}$ = 1120 assuming $G$ = 70 dB, $L_{\mathrm{mix}}$ = -7 dB and $L_{\mathrm{att}}$ = -2 dB.

\subsection*{Noise measurements} For the noise measurements, the measurement setup was modified in the following ways, see Figure 5\textbf{b}. To remove the phase noise between $v_{\mathrm{LO}}$ and $v_{\mathrm{RF}}$, the two signals were obtained from a splitter supplied by a single voltage source. Furthermore, an adjustable attenuator was installed to room temperature to control the excitation voltage $v_{\mathrm{in}}$. The device was operated in so-called persistent current mode, in which the magnetic flux $m\Phi_{0}$ within the loop alone determines the operating point, i.e., $I_{\mathrm{s}} = m\Phi_{0}/(L_{\mathrm{g}}+L_{\mathrm{k}})$, eliminating the noise of the DC flux bias. A signal composed mainly of the Q-component was chosen for the measurement by using cables of proper length acting as a delay line. The device gain was set close to the maximum by biasing the device near the resonance frequency and adjusting the excitation voltage $v_{\mathrm{in}}$ to 16 mV. The voltage noise at the output was measured both in (550 nV/Hz$^{1/2}$) and off (410 nV/Hz$^{1/2}$) the resonance (Figure 6), and we consider the latter as the noise floor of the electronics.  The device gain was checked after and before the noise measurement yielding $G_{\mathrm{tot}}dV/dB$ = 23.7 V/$\mu$T. This gives an equivalent noise $S_{B,\mathrm{sys}}^{1/2}$ = 17 fT/Hz$^{1/2}$ for the system.
\begin{figure}[htpb]
\begin{center}
\includegraphics[width=8cm]{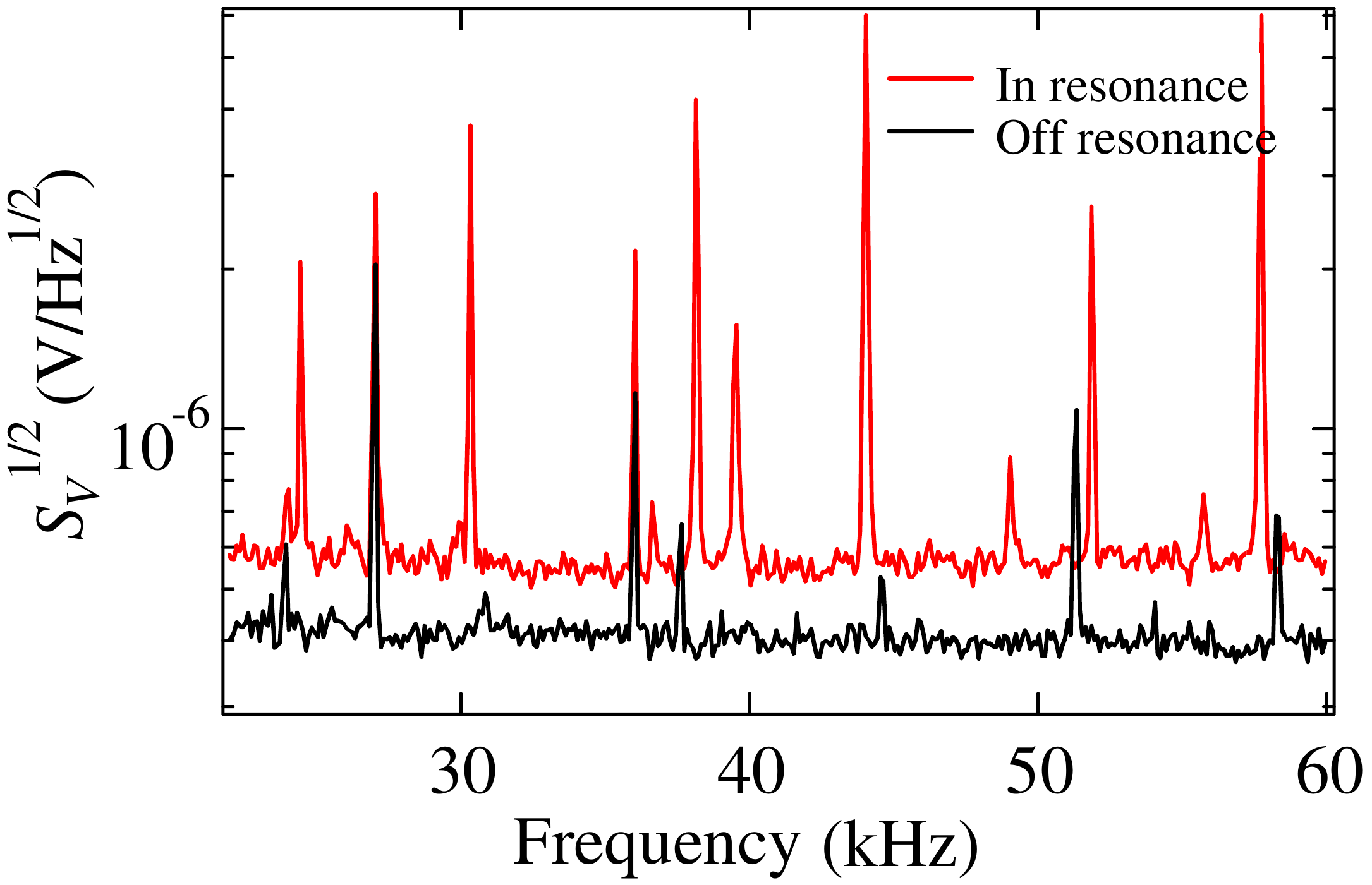}
\caption{Voltage noise $S_{V}^{1/2}$ at the output measured in and off the resonance.}  
\end{center}
\end{figure}

The noise from the readout amplifier chain and the white noise around RF excitation frequency can be formally described as an increase in equivalent noise temperature above the bath temperature. It is common for these noise sources that, when referred to output voltage, they are not affected by the responsivity. In comparison to the theoretical value of Johnson noise, we can estimate the equivalent noise temperature of our system $T_e\approx $ $(S_{B,\mathrm{sys}}/S_{B,\mathrm{th}}$)$\times T$. With $S_{B,\mathrm{th}}^{1/2}$ = 2 fT/Hz$^{1/2}$ and $T$ = 4.2 K, we get $T_e\approx$ 290 K.

The stochastics of the RF excitation may also contribute. To eliminate the effect of phase noise, we used the same frequency source for both the carrier and the local oscillator. In our RF operating point at the bias frequency corresponding to the transmission minimum, the I-component responsivity is zero and the output is to first order insensitive to RF amplitude fluctuations. However, the inductance non-linearity partially rectifies the RF excitation  signal. In Equation (2), the RF excitation current in the magnetometer loop can be contained by additional current term $(i_L/2)\sin(\omega t)$, where the factor of 1/2 stems from the fact that the current is divided into two branches. Thus, generalizing $I_{\mathrm{s}}' = I_{\mathrm{s}} +(i_{L}/2)\sin(\omega t)$, replacing $I_{\mathrm{s}}$ by $I_{\mathrm{s}}'$ in Equation (2) and omiting extra RF terms, we get $L_{\mathrm{k}} = L_{\mathrm{k0}}(1 + (I_{\mathrm{s}}/I^{*})^{2}) + L_{\mathrm{RF}}$ with $L_{\mathrm{RF}} = (L_{\mathrm{k0}}/8)(i_L/I^{*})^2$. If there now exists RF amplitude fluctuation $S_{i_{L}}$, it can be mapped to equivalent magnetic field noise $S_{B,i_{L}}^{1/2}= \left( \partial L/\partial B\right) ^{-1}\left( \partial L_{\mathrm{RF}}/\partial i_{L}\right) S_{iL}^{1/2}$, which is readily expressed as 
\begin{equation}
S_{B,I_L}^{1/2}=\left( \frac{\partial L}{\partial B}\right) ^{-1}\frac{L_{k0}}{4}\left( \frac{I_{L}}{I^{\ast }}\right) ^{2}\frac{S_{iL}^{1/2}}{I_{L}}.
\end{equation}
Now the relative fluctuation $S_{iL}^{1/2}/I_L$ can be traced back to RF source, i.e., $S_{iL}^{1/2}/I_L$ = $S_{v_{\mathrm{RF}}}^{1/2}/v_{0,\mathrm{RF}}$. We measured this with the excitation level used in noise measurements and the result was $S_{v_{\mathrm{RF}}}^{1/2}/v_{0,\mathrm{RF}}$ = 32$\times$10$^{-9}$ Hz$^{-1/2}$. The RF current was $I_{L} \approx$ 4 mA yielding $S_{B,I_L}^{1/2} \approx$ 10 fT/Hz$^{1/2}$.

\subsection*{Generation-recombination noise}We derive the equivalent magnetic field noise $S_{B,\mathrm{gr}}$ stemming from the quasiparticle number fluctuation $S_{N}(\omega)=4n_{\mathrm{qp}}V\tau _{\mathrm{r}}/(1+\left( \omega \tau_{\mathrm{r}}\right) ^{2})$. The kinetic inductance can be estimated as $L_{\mathrm{k}}\approx(n/n_{\mathrm{s}})L_{\mathrm{k0}}\approx(1 + n_{\mathrm{qp}}/n_{\mathrm{s}})L_{k0}=(1 + N_{\mathrm{qp}}/(Vn))L_{\mathrm{k0}}$, where we approximate $n_{\mathrm{s}} \approx n$, i.e., a majority of the charge carriers is paired. The inductance fluctuation is derived  as $S_L (\omega) = (\partial L/\partial N_{\mathrm{qp}})^2 S_{N}(\omega) = (L_{\mathrm{k0}}/Vn_{\mathrm{s}})^2 S_{N}(\omega)$. Equation (7) follows now from $S_{B,\mathrm{gr}}(\omega) = (\partial L/ \partial B)^{-2} S_L (\omega)$ assuming $\omega\ll\tau_{\mathrm{r}}^{-1}$.

Thus, the generation-recombination noise depends on parameters  $n_{\mathrm{s}}$, $n_{\mathrm{qp}}$ and $\tau_{\mathrm{r}}$. As noted above, we assume that the paired carrier density $n_{\mathrm{s}}$ approximately equals the total density $n$ of electrons, i.e., $n_{\mathrm{s}}\approx n$ $\sim 10^{29}$ 1/m$^{3}$ \cite{cho}. For our NbN films, the Cooper pair density is empirically found to follow relation $n_{\mathrm{s}}/(n_{\mathrm{s}}+n_{\mathrm{qp}})$ = $1-(T/T_{\mathrm{C}})^{2.5}$\cite{tim}. Using this combined with the relation of the intrinsic quality factor $Q_{\mathrm{i}} = n_{\mathrm{s}}/(n_{\mathrm{qp}}\omega_{0}\tau_{qp})$, where $\tau_{qp}$ is the quasiparticle thermalization time with a superconductor lattice, and the data shown in Figure 2\textbf{c}, gives $n_{\mathrm{qp}}/n_{\mathrm{s}}$  $\approx$ 0.1. The recombination time can be expressed as \cite{vis}
\begin{equation}
\tau_{\mathrm{r}}=\frac{\tau_{0}}{\sqrt{\pi}}\left(\frac{k_{\mathrm{B}}T_{\mathrm{c}}}{2\Delta}\right)^{5/2}\sqrt{\frac{T_{\mathrm{c}}}{T}}e^{\Delta/k_{\mathrm{B}}T},
\end{equation}
where $\tau_{0}$ is material-dependent electron-phonon interaction time and $\Delta$ is the energy gap of the superconductor. For NbN, an empirical relation $\tau_{0}$[s] = 5$\times10^{-10}$$T$[K]$^{-1.6}$ is typically found valid \cite{gou} leading to $\tau_0\approx$ 50 ps. The zero-temperature energy gap for NbN is $\Delta$ = 2.5 meV \cite{kam}. Although this is likely somewhat suppressed at the operating point, it will give a worst-case estimate for the noise. Finally, we get $\tau_{\mathrm{r}} \approx$ 1.5 ns and $S_{B,\mathrm{gr}}^{1/2} \approx$ 0.03 fT/Hz$^{1/2}$ using $dL/dB \approx$ 2 mH/T at operating point.

\subsection*{Calibration of mutual inductances} A square-shaped two-turn PCB coil (inductance $L_{1}$) and an on-chip superconducting coil (inductance $L_{2}$) were arranged around the sample to introduce magnetic bias and calibration fields. The mutual inductance $M_{1}$ between the magnetometer and the off-chip coil was evaluated using the analytic formula of magnetic field around a straight wire of finite length
\begin{equation}
	B=\frac{\mu_{0}I}{4\pi r}\left(\cos \theta_{1}+\cos \theta_{2}\right),
\end{equation}
where the current $I$ flows from one end (point A) of the wire to the other (point B) and the magnetic field is calculated a distance $r$ away from the wire (point C) defined by the angles $\theta_{1}$ = $\angle$(ABC) and $\theta_{2}$ = $\angle$(CAB). The model was found to reproduce the field profile measured with a magnetic field sensor (Alphalab Gaussmeter GM2) approximately 1 mm above the PCB yielding $M_{1}$ = 49$\pm$3 nH, where the error limit is related to the uncertainty in the position of the magnetometer. In addition, the calculated $M_{1}$, combined with the derived inductances $L_{\mathrm{g}}$ and $L_{\mathrm{k0}}$, gives $I_{c}$ = 10 mA for the magnetometer loop in line with the critical current measurements, verifying the validity of the computation method. Mutual inductance $M_{2}$ was calibrated against $M_{1}$ by biasing the magnetometer in the Earth's magnetic field and recording the signal at the output, the probe signal injected either through $M_{1}$ or $M_{2}$. The result was $k$ = $M_{2}/M_{1}$ = 3.2$\pm$0.1 leading to $M_{2}$ = 157$\pm$14 nH. Again, the calibration seems to be accurate as discovered from responsitivity measurements with $G_{\mathrm{tot}}\approx G_{\mathrm{nom}}$, see above.

\end{document}